\documentclass[twocolumn,prl,preprintnumbers,tightenlines,footinbib,superscriptaddress]{revtex4}

\pdfoutput=1
\usepackage{amsmath,amssymb,amsfonts}
\usepackage{graphicx}
\usepackage[normalem]{ulem}
\usepackage{slashed}
\usepackage{booktabs}
\usepackage{hyperref}
\usepackage{color}
\usepackage[sort&compress]{natbib}
\usepackage{tikz}
\usepackage{cleveref}
\newcounter{qnumber}

\newcommand{\drawsquare}[2]{\hbox{%
\rule{#2pt}{#1pt}\hskip-#2pt
\rule{#1pt}{#2pt}\hskip-#1pt
\rule[#1pt]{#1pt}{#2pt}}\rule[#1pt]{#2pt}{#2pt}\hskip-#2pt
\rule{#2pt}{#1pt}}

\newcommand{\Yfund}{\raisebox{-.5pt}{\drawsquare{6.5}{0.4}}}
\newcommand{\Ysymm}{\raisebox{-.5pt}{\drawsquare{6.5}{0.4}}\hskip-0.4pt%
        \raisebox{-.5pt}{\drawsquare{6.5}{0.4}}}
\newcommand{\Yasymm}{\raisebox{-3.5pt}{\drawsquare{6.5}{0.4}}\hskip-6.9pt%
        \raisebox{3pt}{\drawsquare{6.5}{0.4}}}

\begin{document}

\title{Demonstration of Confinement and Chiral Symmetry Breaking\\in $SO(N_c)$ Gauge Theories}

\author{Csaba Cs\'aki}
\email{ccsaki@gmail.com}
\affiliation{Department of Physics, LEPP, Cornell University, Ithaca, NY 14853, USA}

\author{Andrew Gomes}
\email{awg76@cornell.edu}
\affiliation{Department of Physics, LEPP, Cornell University, Ithaca, NY 14853, USA}

\author{Hitoshi Murayama}
\email{hitoshi@berkeley.edu, hitoshi.murayama@ipmu.jp, Hamamatsu Professor}
\affiliation{Department of Physics, University of California, Berkeley, CA 94720, USA}
\affiliation{Kavli Institute for the Physics and Mathematics of the
  Universe (WPI), University of Tokyo,
  Kashiwa 277-8583, Japan}
\affiliation{Ernest Orlando Lawrence Berkeley National Laboratory, Berkeley, CA 94720, USA}

\author{Ofri Telem}
\email{t10ofrit@gmail.com}
\affiliation{Department of Physics, University of California, Berkeley, CA 94720, USA}
\affiliation{Ernest Orlando Lawrence Berkeley National Laboratory, Berkeley, CA 94720, USA}

\begin{abstract} 
We demonstrate that $SO(N_{c})$ gauge theories with matter fields in the vector representation confine due to monopole condensation and break the $SU(N_{F})$ chiral symmetry to $SO(N_{F})$ via the quark bilinear. Our results are obtained by perturbing the ${\cal N}=1$ supersymmetric theory with anomaly-mediated supersymmetry breaking.
\end{abstract}

\maketitle

\section{Introduction}

Ever since quarks were proposed as fundamental constituents of the proton, neutron, and numerous hadrons by Gell-mann and Ne'eman \cite{GellMann1961,Neeman1961}, it has been a mystery why they cannot be observed directly in experiments. At the same time, protons and neutrons bind in atomic nuclei due to the exchange of light pions predicted by Yukawa \cite{Yukawa1935}. The binding of nuclei, and correspondingly the entire world of chemistry, hinges on pions being much lighter than protons, despite the fact that they are made of the same quarks. The first mystery was ``explained'' by postulating {\it confinement}\/ of quarks by condensation of magnetic monopoles via the dual Mei\ss ner effect proposed by Mandelstam \cite{Mandelstam1976} and 't~Hooft \cite{tHooft:1981bkw}. The second mystery was ``explained'' by postulating {\it chiral symmetry breaking}\/ whose the Nambu--Goldstone bosons are the light pions proposed by Nambu and Jona-Lasinio \cite{Nambu:1961tp,Nambu:1961fr}. In either case, however, it has been a challenge to derive these properties from the fundamental theory of strong interactions, Quantum ChromoDynamics (QCD).

It has been proposed recently \cite{Murayama:2021xfj} that one can study the dynamics of gauge theories using the supersymmetric version of the theory perturbed in a specific way called anomaly-mediated supersymmetry breaking (AMSB) \cite{Randall:1998uk,Giudice:1998xp} (see also \cite{ArkaniHamed:1998kj,Arkani-Hamed:1998dti} for earlier work containing some important aspects of AMSB). For other analyses of non-supersymmetric gauge theories via controlled supersymmetry breaking, see, for example, \cite{Evans:1995rv,Evans:1995ia,Aharony:1995zh,Alvarez-Gaume:1996qoj,Alvarez-Gaume:1996vlf,DHoker:1996xdz,Evans:1996hi,Konishi:1996iz,Alvarez-Gaume:1997bzm,Alvarez-Gaume:1997wnu,Evans:1997dz,Cheng:1998xg,Martin:1998yr,Luty:1999qc}, as well as the more recent \cite{Cordova:2018acb}. When AMSB was applied to $SU(N_{c})$ QCD, it was possible to derive chiral symmetry breaking for $1 < N_{F} \leq \frac{3}{2} N_{c}$, while the theory flows to a conformal fixed point for $\frac{3}{2}N_{c} < N_{F} \leq 3 N_{c}$. Yet the $SU(N_{c})$ theory does not confine in the presence of quarks in the fundamental representation because any color charges can be screened.

The $SO(N_{c})$ theory with fermions in the vector representation is interesting because it does truly confine, since the spinor representation transforming under the ${\mathbb Z}_{2}$ center cannot be screened. Therefore, we can hope to see the interplay between the condensation of monopoles on one hand, and fermion bilinears on the other hand. It turns out that we should focus on $N_{F} \leq N_{c}-2$ where we can demonstrate monopole condensation.

In this Letter, we sketch the essence of the analysis, while details are presented in a forthcoming companion paper \cite{companion}, that will also contain a discussion of the cases where $N_{F} > N_{c}-2$.

\section{Anomaly Mediation}

Anomaly mediation of supersymmetry breaking (AMSB) is parameterized by a single number $m$ that explicitly breaks supersymmetry in two different ways. One is the tree-level contribution based on the superpotential
\begin{align}
	{V}_{\rm tree} &= m \left( \varphi_{i} \frac{\partial W}{\partial \varphi_{i}} - 3 W \right)
	+ c.c. \label{eq:AMSBW}
\end{align}
Note that Eq.~(\ref{eq:AMSBW}) also breaks the $U(1)_R$ symmetry explicitly.
When the superpotential does not include dimensionful parameters, this expression identically vanishes. In this case, there are the loop-level supersymmetry breaking effects from the superconformal anomaly \cite{Pomarol:1999ie}. In this Letter, we do not need the loop-level effects that can be neglected in the presence of the tree-level effects \eqref{eq:AMSBW}. The loop-level effects will be discussed in the companion paper \cite{companion} for special cases when they are needed.


\section{$N_F=N_c-2$}
We consider an $SO(N_c)$ gauge theory with $N_F=N_c-2$ flavors $Q^i$. In the supersymmetric limit, the theory is in an abelian Coulomb phase \cite{Intriligator:1995id}. The $D$-flat directions are parameterized by the diagonal entries of the mesons $M^{ij}\equiv Q^iQ^j$. As $M^{ij}$ are neutral under $U(1)_R$, no superpotential can be generated, and there is a quantum moduli space. At a generic point $M^{ij}$ on the meson moduli space, the gauge symmetry is higgsed to a $U(1)$, and so the theory only has a Coulomb branch. The effective gauge coupling $\tau=\frac{\theta}{2\pi}+\frac{i8\pi}{g^2}$ of the theory is given on the Coulomb branch as a function of the $SU(N_F)$ invariant $U\equiv\text{det}M$ only. There are singularities at the two points $U=U_1 \equiv 16\Lambda^{2N_F}$ and $U=0$. 

Around the singular point $U=U_1$, the relevant light degrees of freedom are the monopoles $E^{\pm}$ with magnetic charges $\pm1$, which transform under the UV global symmetry $SU(N_F)\times U(1)_R$ as $E^{\pm}(\bf1)_1$. Since $\text{det}M\equiv U\neq0$, the global symmetry at this point is broken to $SO(N_F)\times U(1)_R$. The theory has a dynamically generated superpotential about $U=U_1$ of
\begin{align}
W_{\text{mon}}&=\tilde{f}\left(U-U_1\right)\,E^+\,E^-\,,
\end{align}
where $\tilde{f}(t)=t+\dotsb$ is a holomorphic function in the neighborhood of $t=0$. In practice, only the leading order in $\tilde{f}$ matters for the stabilization of the minimum. Using canonically normalized fields we have
\begin{align}
W_{\text{mon}}&=\Lambda\left(\frac{\tilde{U}}{\Lambda^{N_F}}-16\right)\,\tilde{E}^+\,\tilde{E}^-\,,
\end{align}
where $\tilde{U}=\text{det}\tilde{M}$ and $\tilde{M}=M/\Lambda,\,\tilde{E}^\pm=E^\pm/\sqrt{\Lambda}$ are the canonically normalized meson and monopoles, respectively. Exactly at $\tilde{U}=\tilde{U}_1\equiv16\Lambda^{N_F}$, 't Hooft anomaly matching is saturated by $\tilde{E}^{\pm},\,\tilde{M}^{ij}$, and the photinos $\mathcal{W}_\alpha\sim W_\alpha Q^{N_c-2}$, whose charges are given in Table~\ref{tab:Ncm2U1}. It is easy to verify that the $U(1)_R\,{\rm gravity}^{2}$, $U(1)^{3}_R$, and $U(1)_R\,SO(N_F)^{2}$ anomalies all match. Therefore, we know the degrees of freedom in the IR and their K\"ahler potentials are regular at this singularity.

\begin{table}[t]
	\centerline{
	\begin{tabular}{|c|c|c|c||c|c|} \hline
	& $SO(N_c)$ & $SU(N_F)$&$U(1)_R$&$U(1)_{\text{mag}}$&$SO(N_F)$ \\ \hline
	$Q^i$ & $\Yfund$ & $\Yfund$&$0$&$-$&$\Yfund$ \\ \hline 
	$\lambda$&$\Yasymm$&$\bf 1$&$1$&$-$&$\bf 1$\\ \hline\hline
	 $M^{ij}$&$\bf1$&$\Ysymm$&$0$&$-$&$\bf1+\Ysymm$\\ \hline
	 $E^{\pm}$&$-$&$\bf 1$&$1$&$\pm 1$&$\bf 1$\\ \hline
	 $\lambda_{\text{mag}}$&$-$&$\bf 1$&$1$&$0$&$\bf 1$\\ \hline
	\end{tabular}
	}
	\caption{Degrees of freedom in the $SO(N_c)$ theory with $N_F=N_c-2$ near $U=U_1$. The unbroken global symmetry with $M^{ij}\propto\delta^{ij}$ is $SO(N_F)\times U(1)_R$.}\label{tab:Ncm2U1}
\end{table}

AMSB generates a tree-level contribution to the scalar potential from \eqref{eq:AMSBW}, producing the global minimum at $\tilde{U}= \tilde{U}_1$. In particular, the scalar potential along $\tilde{M}^{ij}=\tilde{M}\delta^{ij}$ is given locally as
\begin{align}\label{eq:Ncm2pot}
\lefteqn{
V_{\tilde{U}\sim \tilde{U}_1}=\Lambda^2\left|{\left(\frac{\tilde{M}}{\Lambda}\right)}^{N_F}-16\right|^2\left(|\tilde{E}^+|^2+|\tilde{E}^-|^2\right) } \nonumber\\
&+\frac{1}{kN_F}\left|N_F{\left(\frac{\tilde{M}}{\Lambda}\right)}^{N_F-1}\right|^2|\tilde{E}^+\tilde{E}^-|^2\,+\,V_{\text{AMSB}}\,.
\end{align}
Note the $(k N_F)^{-1}$ factor in the second line, which comes from the K\"ahler term $k N_F \tilde{M}^\dagger \tilde{M}$ for $\tilde{M}$, where $k$ is an unknown $\mathcal{O}(1)$ normalization factor. The tree-level AMSB contribution is given by \eqref{eq:AMSBW}, {\it i.e.}\/,
\begin{eqnarray}\label{eq:Ncm2AMSB}
V_{\text{AMSB}}=m\Lambda\left[16+(N_F-1){\left(\frac{\tilde{M}}{\Lambda}\right)}^{N_F}\right]\tilde{E}^+\tilde{E}^-+\text{c.c.}\,
\end{eqnarray}
This potential has a minimum at
\begin{eqnarray}\label{eq:globalNcm2}
&&\tilde{M}=16^{\frac{1}{N_F}}\Lambda,
\qquad |\tilde{E}^+||\tilde{E}^-|=16^{\frac{2}{N_F}-1}\,km\Lambda,\nonumber\\[5pt]
&&~~V_{\text{min}}=-16^{\frac{2}{N_F}}N_F\,km^2\Lambda^2\,.
\end{eqnarray}
Since $\tilde{M}^{ij}=\tilde{M}\delta^{ij}$ in this minimum, the global symmetry is broken spontaneously to $SO(N_f)$, while $U(1)_R$ is explicitly broken by AMSB, and there are no 't Hooft anomalies to match. 

The most remarkable feature of the minimum \eqref{eq:globalNcm2} is the condensation of monopoles $\tilde{E}^\pm$, which gives an area law to non-trivial Wilson loop operators, indicating confinement \cite{tHooft:1981bkw,Cardy:1981qy,Cardy1982}. This phenomenon is well known in the context of the breaking of $\mathcal{N}=2$ Seiberg-Witten theory to $\mathcal{N}=1$ by introducing a tree level superpotential for the matter field \cite{Seiberg:1994rs}. In \cite{Evans:1996hi,Konishi:1996iz}, monopole condenstaion was shown in a non-supersymmetric theory by introducing soft SUSY breaking on top of the superpotential term for the Seiberg-Witten model. Here, monopole condensation and SUSY breaking emerge together as a result of AMSB. Furthermore, since the global $SU(N_F)$ symmetry is broken to $SO(N_F)$, this is an example of \textit{confinement with chiral symmetry breaking} in a non-supersymmetric theory. 

In the large $m$ limit where all scalar superpartners decouple, we can connect the chiral symmetry breaking observed here to the familiar one due to fermion bilinears. To see this, note that in the large $m$ limit the fermion bilinears are identified with the $F$-component of the meson chiral superfield:
\begin{align}
\langle \psi_{i}^* \psi_{j}^* \rangle = F^*_{M_{ij}} = 16 \Lambda^2 M^{-1}_{ij} E^{+} E^{-} \propto \delta_{ij} k m \Lambda^{2} \neq 0.
\end{align}
In other words, our analysis demonstrates the condensation of fermion bilinears in a non-supersymmetric theory, in addition to the monopole condensate.

Around the singular point $U=0$ the relevant light degrees of freedom are the dyons $q^{\pm}_i$ with magnetic charge $\pm1$, which transform under the UV global symmetry $SU(N_F)\times U(1)_R$ as $q^{\pm}_i(\overline{\Yfund})_1$.
These have a dynamically generated superpotential about $U=0$ of
\begin{align}\label{eq:monYuk}
W_{\text{dyon}}&=\frac{1}{\mu}\,f(t)\,M^{ij}q^+_{i}q^-_{j}\,,
\end{align}
where $\mu$ is an effective mass scale, $t=U\Lambda^{4-2N_c}$, and $f(t)$ is a holomorphic function in the neighborhood of $t=0$, normalized so that $f(0)=1$. However, the scale $\mu$ can be absorbed into the normalization of the meson field $\hat{M}=M/\mu$ and the theory at this point has no dimensionful parameters. Therefore the AMSB is loop-suppressed, and hence so is the vacuum energy. Consequently, the local AMSB minimum near this singularity is \textit{not} the global minimum.

\begin{table*}[t]
	\centerline{
	\begin{tabular}{|c|c|c|} \hline
	Range& SUSY &+AMSB \\ \hline
	$N_F=1$ & run-away  & confinement \\ \hline
	$1<N_F<N_c-4$ & run-away &  confinement$+ \chi$SB \\ \hline
	$N_F=N_c-4$ & 2 branches & confinement$+ \chi$SB \\ \hline
	$N_F=N_c-3$ & 2 branches & confinement$+ \chi$SB\\ \hline
	$N_F=N_c-2$ & abelian Coulomb & confinement$+ \chi$SB \\ \hline
	$N_F=N_c-1$ & free magnetic, 2 branches & confinement$+ \chi$SB\\ \hline
	$N_F=N_c$ & free magnetic, 2 branches & confinement$+ \chi$SB\\ \hline
	$N_c+1\leq N_F\leq \frac{3}{2}(N_c-2)$& free magnetic &  confinement$+ \chi$SB\\ \hline
	$\,\frac{3}{2}(N_c-2)<N_F\leq 3(N_c-2)\,$ & CFT & CFT \\ \hline
	$3(N_c-2)<N_F$& IR free & run-away \\ \hline
	\end{tabular}
	}
	\caption{Summary of the IR behavior of $SO(N_c)$ theories with $N_F$ fundamentals with AMSB. $\chi$SB stands for chiral symmetry breaking. For $N_F=N_c-1$ and $N_c$, two branches appear along the flat direction of the maximum rank of the meson $M^{ij}$, yet the AMSB chooses one over the other, resulting in the $\chi$SB.}\label{tab:summary}
\end{table*}

\section{Monopole Condensation for $N_F<N_c-2$ via Mass Deformations}

In the above discussion of the theory with $N_F=N_c-2$, we explicitly saw monopole condensation in the non-supersymmetric vacuum of the theory. Here we wish to study the cases with fewer flavors, by treating the latter as the $N_F=N_c-2$ deformed by mass terms $\mu$, with $\mu\gg\Lambda$. In this way, we will be able to interpret the theories with fewer flavors as also corresponding to monopole condensation all the way down to the pure $SO(N_c)$ Yang--Mills case. On the other hand, we can also study the same theory with the Affleck--Dine--Seiberg (ADS) superpotential perturbed by AMSB. They must agree if we believe in the holomorphy argument that $\mu$, $m$, and $\Lambda$ can be varied without a phase transition. 

We begin by considering the $N_F=N_c-2$ theory in the supersymmetric limit, with just one mass term for the last flavor,
\begin{align}\label{eq:Ncm2mass}
W&=\Lambda\left(\frac{\det \tilde{M}}{\Lambda^{N_F}}-16\right)\,\tilde{E}^+\,\tilde{E}^- + \frac{1}{2}\mu \Lambda\tilde{M}^{N_F N_F}
\end{align}
The equation of motion for $\tilde{M}^{N_F N_F}$ gives
\begin{align}\label{eq:EElower}
\tilde{E}^+ \tilde{E}^- &= -\frac{1}{2} \frac{\mu \Lambda^{N_F}}{\det \tilde{M}'}\, ,
\end{align}
where $\tilde{M}'$ is the matrix of the remaining mesons. 

On the other hand, the extra flavor can be integrated out first to give the ADS superpotential
\begin{align}\label{eq:ADS}
W_{\text{ADS}}&=\frac{N_c-N^\prime_F-2}{2}\omega^k\left(\frac{16\Lambda^{\prime 3N_c-N^\prime_{F}-6}}{\text{det} \tilde{M}^\prime}\right)^{\frac{1}{ N_{c}-N^\prime_{F}-2}}\,,
\end{align}
where $N^\prime_{F}=N_{F}-1=N_c-3$, and $\Lambda^{\prime 3N_{c}-N_{F}-5} = \mu \Lambda^{3N_{c}-N_{F}-6}$ is the strong scale of the theory and $\omega = e^{2\pi i/(N_c-N^\prime_F-2)}$ with $k=0, 1, \dots, N_c-N^\prime_F-3$. Since $N^\prime_{F} = N_{c}-3$, there is another branch on which the superpotential vanishes; we have checked that this branch does not produce the global minimum when turning on AMSB. The SUSY theory runs away and does not have a ground state. Turning on AMSB stabilizes the runaway behavior of the superpotential at a large amplitude where the K\"ahler potential is canonical for $\varphi \gg \Lambda$ with $M^{ij}=\varphi^2\delta^{ij}$. The tree-level AMSB is
\begin{equation}\label{eq:NFlNcm4AMSB}
V_{\text{AMSB}}=-m\Lambda^{\prime 3}\,\frac{3N_c-N^\prime_F-6}{2}\left(\frac{16\Lambda^{2N^\prime_{F}}}{\varphi^{2N^\prime_F}}\right)^{\frac{1}{ N_{c}-N^\prime_{F}-2}}+\text{c.c.}\,,
\end{equation}
which together with the scalar potential derived from the superpotential \eqref{eq:ADS} gives a minimum
\begin{align}\label{eq:globalNFlNcm4}
\varphi&=2^{\frac{2}{N_c-2}}\,{\left(f_{N^\prime_F}\frac{\Lambda'}{m}\right)}^{\frac{N_c-N^\prime_F-2}{2(N_c-2)}}~\Lambda' \nonumber\\
V_{\text{min}}&=-2^{\frac{4}{N_c-2}}\frac{N_c-2}{f^2_{N^\prime_F}}{\left(f_{N^\prime_F}\frac{\Lambda'}{m}\right)}^{\frac{N_c-N^\prime_F-2}{N_c-2}}~m^2\Lambda^{\prime 2}\,, 
\end{align}
with $f_{N^\prime_F}=\frac{N_c+N^\prime_F-2}{3N_c-N^\prime_F-6}$.  

\begin{figure}[t]
\begin{center}
\includegraphics[width=\columnwidth]{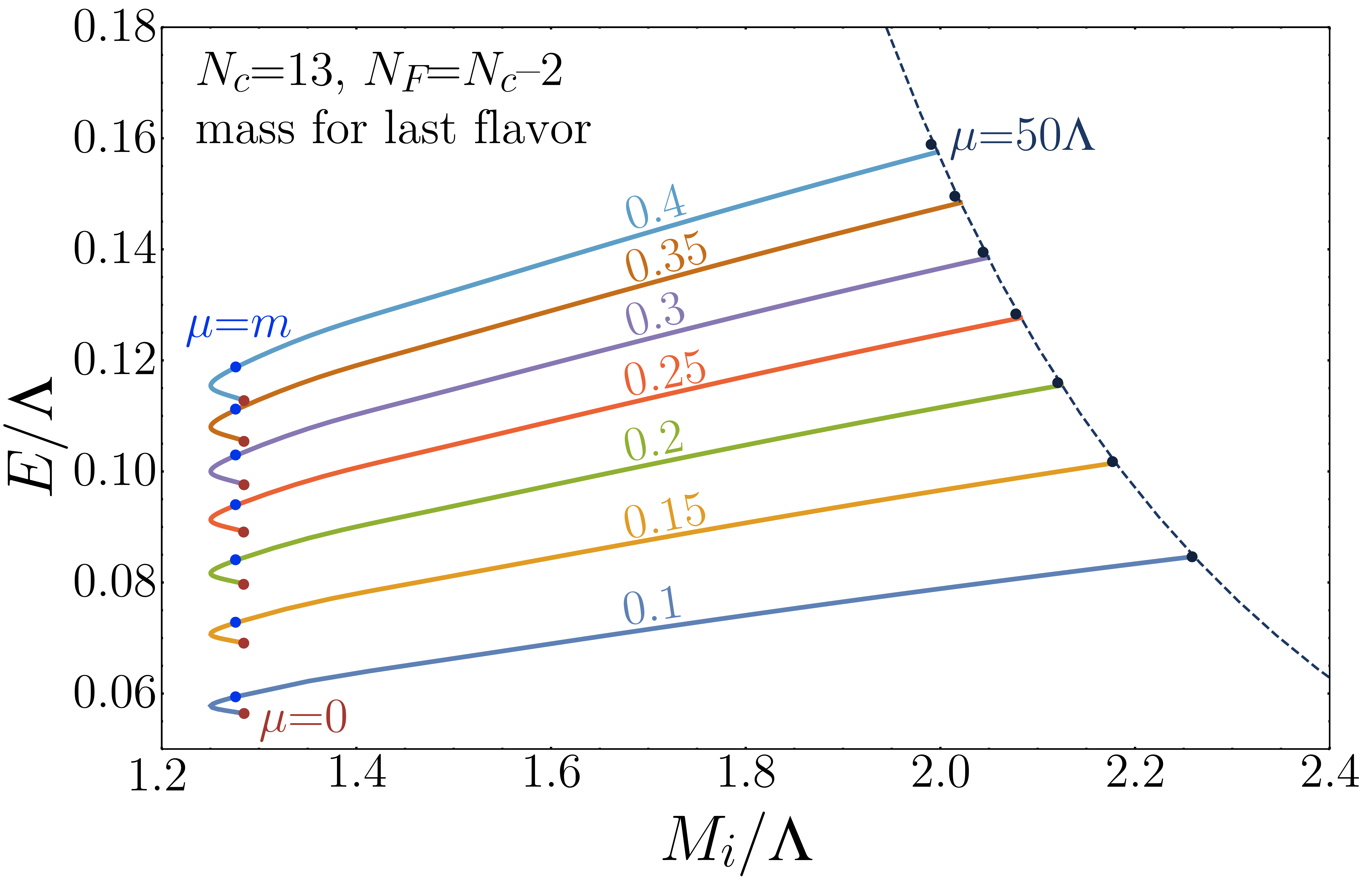}
\caption{Location of the minimum in the theory with AMSB and $N_F=N_c-2$, deformed by a mass term $\frac{1}{2}\mu M^{N_FN_F}$ for the last flavor. $E$ is the VEV of the condensed monopoles, while $M_{i}$ is the common VEV of the first $N_c-3$ flavors. The different curves are labeled by the value of $m/\Lambda$. The curves start at the $N_c-2$ minimum \eqref{eq:globalNcm2} for $\mu=0$.  The VEV of $M_i$ initially decreases by a very small amount but then increases as $\mu$ crosses $m$. As $\mu/\Lambda\rightarrow \infty$, the minimum goes over to the one given by eq. \eqref{eq:globalNFlNcm4} with $N^\prime_F=N_c-3$, while the VEV for the monopoles $E$ persists. The dashed line for $\mu=50\Lambda$ is the relation \eqref{eq:EElower}. Note that the endpoints are a tiny bit off the dashed line as a result of the interpolating K\"ahler potential \eqref{eq:Kinterp}. We have chosen  $N_c=13$ for this plot.}\label{fig:mudef1}
\end{center}
\end{figure}

In Fig.~\ref{fig:mudef1} we show the minimum of the mass-deformed theory ~\eqref{eq:Ncm2mass} in the presence of AMSB. As can be seen in the plot, the VEV of the first $N_c-3$ flavors interpolates between the minimum \eqref{eq:globalNcm2} for $\mu=0$, and the ADS+AMSB minimum \eqref{eq:globalNFlNcm4} with $N^\prime_F=N_c-3$ and $\Lambda\rightarrow\Lambda'$ in the large $\mu$ limit. We can see that the monopole condensate persists in the large $\mu$ limit. 

To correctly reproduce the ADS+AMSB minimum, we had to interpolate the K\"ahler potential between the neighborhood of $\text{det} \tilde{M} \sim \tilde{U}_1$, where it is canonical in $\tilde{M}$, to large $\text{det} \tilde{M}$, where the K\"ahler potential is canonical in $\varphi\sim\sqrt{\tilde{M}\Lambda}$. More specifically, we used the following interpolating K\"ahler potential in the numerical study:
\begin{eqnarray}\label{eq:Kinterp}
K_{\text{interp.}}=\Lambda^2\,\sqrt{1+\frac{\tilde{M}\tilde{M}^\dagger}{\Lambda^2}}\,.
\end{eqnarray}
Interestingly, for $\mu<m$, the UV theory itself is unstable, and has a runaway at $\tilde{E}^+\tilde{E}^-=0$ and $\tilde{M}_i\rightarrow\infty$. This is a feature of the mass term in \eqref{eq:Ncm2mass} in the presence of AMSB, and is unrelated to the dynamics of the gauge theory. Since this does not affect our analysis, we follow the local minimum which is continuously connected to the global minimum for $\mu>m$. This accounts for the small ``U-turn" of the curves in Fig.~\ref{fig:mudef1} between the red points ($\mu=0$) and the blue points ($\mu=m$). Note that our argument regarding monopole condensation in the large $\mu$ limit is completely free of this subtlety.

We explicitly checked that the same conclusions hold when integrating out more than one flavor, such that $0\leq N^\prime_F\leq N_c-2$. Similarly to the $N^\prime_F=N_c-3$ case, for $N^\prime_F=N_c-4$ we have another branch with vanishing superpotential, which does not produce the global minimum in the presence of AMSB.
For all $N^\prime_F$ in the range $0\leq N^\prime_F\leq N_c-2$, we find that monopole condensation persists in the AMSB global minimum. Since in the $\mu\rightarrow \infty$ limit, all of the extra flavors effectively decouple, this is a demonstration of monopole condensation for the entire range $0\leq N_F\leq N_c-2$. 

\section{Larger $N_{F}$}

For $N_{c} - 2 < N_{F} \leq \frac{3}{2} (N_{c}-2)$, the SUSY limit has the IR description in terms of the free magnetic $SO(N_{F}-N_{c}+4)$ theory with magnetic quarks and mesons. With AMSB, the global minimum is obtained when the meson matrix has full rank and the magnetic quarks are integrated out, similar to the case of $SU(N_c)$ QCD \cite{Murayama:2021xfj}. The low-energy limit is a pure $SO(N_{F}-N_{c}+4)$ SUSY Yang--Mills (SYM) with the gaugino condensate, which is known to confine. With AMSB, the fermion biliniear also acquires a VEV, breaking the $SU(N_{F})$ global symmetry to $SO(N_{F})$.

On the other hand for $\frac{3}{2}N_{c} < N_{F} \leq 3 N_{c}$, the SUSY theory flows to conformal fixed point. AMSB effects disappear by a power law towards the fixed point and the theory recovers supersymmetry. 

All these phases are summarized in Table~\ref{tab:summary}, which are discussed in much more detail in the companion paper \cite{companion}.

\section{Conclusions}

We studied the dynamics of the $SO(N_{c})$ gauge theory with fermions in the vector representation using its supersymmetric version perturbed by anomaly mediated supersymmetry breaking. We obtained the exact global minimum and demonstrated that the magnetic monopole and fermion bilinear condense for $N_{F} \leq N_{c}-2$, leading to both confinement and chiral symmetry breaking. We see no indication that there are phase transitions when varying $m$, $\mu$, and $\Lambda$, consistent with the holomorphy argument. This encouraging observation supports the case that our results may persist in the non-supersymmetric limit $m \gg \Lambda$. To the best of our knowledge, this is the first analytic demonstration of both confinement and chiral symmetry breaking in non-supersymmetric gauge theories.

\section{Acknowledgments}

\begin{acknowledgments}

CC is supported in part by the NSF grant PHY-2014071 as well as the US-Israeli BSF grant 2016153. AG is supported in part by the NSERC PGS-D fellowship, and in part by the NSF grant PHY-2014071. OT and HM were supported in part by the DOE under grant DE-AC02-05CH11231.
HM was also supported in part by the NSF grant
PHY-1915314, by the JSPS Grant-in-Aid for
Scientific Research JP20K03942, MEXT Grant-in-Aid for Transformative Research Areas (A)
JP20H05850, JP20A203, by WPI, MEXT, Japan, and Hamamatsu Photonics, K.K.

\end{acknowledgments}

\bibliographystyle{utcaps_mod}
\bibliography{AS_SO}

\providecommand{\href}[2]{#2}\begingroup\raggedright\begin{thebibliography}{10}

\bibitem{GellMann1961}
M.~Gell-Mann, ``{\em The Eightfold Way: A Theory of Strong Interaction
  Symmetry},''. \url{https://www.osti.gov/biblio/4008239}.

\bibitem{Neeman1961}
Y.~Ne'eman, ``{\em Derivation of strong interactions from a gauge
  invariance},''
  \href{http://dx.doi.org/https://doi.org/10.1016/0029-5582(61)90134-1}{Nuclear
  Physics {\normalfont \bfseries 26} (1961) no.~2, 222--229}.

\bibitem{Yukawa1935}
H.~Yukawa, ``{\em {On the Interaction of Elementary Particles. I}},''
  \href{http://dx.doi.org/10.1143/PTPS.1.1}{Progress of Theoretical Physics
  Supplement {\normalfont \bfseries 1} (1955)  1--10}.

\bibitem{Mandelstam1976}
S.~Mandelstam, ``{\em Vortices and quark confinement in non-abelian gauge
  theories},''
  \href{http://dx.doi.org/https://doi.org/10.1016/0370-2693(75)90221-X}{Physics
  Letters B {\normalfont \bfseries 53} (1975) no.~5, 476--478}.

\bibitem{tHooft:1981bkw}
G.~'t~Hooft, ``{\em {Topology of the Gauge Condition and New Confinement Phases
  in Nonabelian Gauge Theories}},''
  \href{http://dx.doi.org/10.1016/0550-3213(81)90442-9}{Nucl. Phys. B
  {\normalfont \bfseries 190} (1981)  455--478}.

\bibitem{Nambu:1961tp}
Y.~Nambu and G.~Jona-Lasinio, ``{\em {Dynamical Model of Elementary Particles
  Based on an Analogy with Superconductivity. I}},''
  \href{http://dx.doi.org/10.1103/PhysRev.122.345}{Phys. Rev. {\normalfont
  \bfseries 122} (1961)  345--358}.

\bibitem{Nambu:1961fr}
Y.~Nambu and G.~Jona-Lasinio, ``{\em {Dynamical Model of Elementary Particles
  Based on an Analogy with Superconductivity. II}},''
  \href{http://dx.doi.org/10.1103/PhysRev.124.246}{Phys. Rev. {\normalfont
  \bfseries 124} (1961)  246--254}.

\bibitem{Murayama:2021xfj}
H.~Murayama, ``{\em {Some Exact Results in QCD-like Theories}},'' Phys. Rev.
  Lett. (2021)  in print, \href{http://arxiv.org/abs/2104.01179}{{\normalfont
  \ttfamily arXiv:2104.01179}}.

\bibitem{Randall:1998uk}
L.~Randall and R.~Sundrum, ``{\em {Out of this world supersymmetry
  breaking}},'' \href{http://dx.doi.org/10.1016/S0550-3213(99)00359-4}{Nucl.
  Phys. B {\normalfont \bfseries 557} (1999)  79--118},
  \href{http://arxiv.org/abs/hep-th/9810155}{{\normalfont \ttfamily
  arXiv:hep-th/9810155}}.

\bibitem{Giudice:1998xp}
G.~F. Giudice, M.~A. Luty, H.~Murayama, and R.~Rattazzi, ``{\em {Gaugino mass
  without singlets}},''
  \href{http://dx.doi.org/10.1088/1126-6708/1998/12/027}{JHEP {\normalfont
  \bfseries 12} (1998)  027},
  \href{http://arxiv.org/abs/hep-ph/9810442}{{\normalfont \ttfamily
  arXiv:hep-ph/9810442}}.

\bibitem{ArkaniHamed:1998kj}
N.~Arkani-Hamed, G.~F. Giudice, M.~A. Luty, and R.~Rattazzi, ``{\em
  {Supersymmetry breaking loops from analytic continuation into superspace}},''
  \href{http://dx.doi.org/10.1103/PhysRevD.58.115005}{Phys. Rev. {\normalfont
  \bfseries D58} (1998)  115005},
\href{http://arxiv.org/abs/hep-ph/9803290}{{\normalfont \ttfamily
  arXiv:hep-ph/9803290}}.

\bibitem{Arkani-Hamed:1998dti}
N.~Arkani-Hamed and R.~Rattazzi, ``{\em {Exact results for nonholomorphic
  masses in softly broken supersymmetric gauge theories}},''
  \href{http://dx.doi.org/10.1016/S0370-2693(99)00406-2}{Phys. Lett. B
  {\normalfont \bfseries 454} (1999)  290--296},
  \href{http://arxiv.org/abs/hep-th/9804068}{{\normalfont \ttfamily
  arXiv:hep-th/9804068}}.

\bibitem{Evans:1995rv}
N.~J. Evans, S.~D.~H. Hsu, M.~Schwetz, and S.~B. Selipsky, ``{\em {Exact
  results and soft breaking masses in supersymmetric gauge theory}},''
  \href{http://dx.doi.org/10.1016/0550-3213(95)00516-5}{Nucl. Phys. B
  {\normalfont \bfseries 456} (1995)  205--218},
  \href{http://arxiv.org/abs/hep-th/9508002}{{\normalfont \ttfamily
  arXiv:hep-th/9508002}}.

\bibitem{Evans:1995ia}
N.~J. Evans, S.~D.~H. Hsu, and M.~Schwetz, ``{\em {Exact results in softly
  broken supersymmetric models}},''
  \href{http://dx.doi.org/10.1016/0370-2693(95)00778-J}{Phys. Lett. B
  {\normalfont \bfseries 355} (1995)  475--480},
  \href{http://arxiv.org/abs/hep-th/9503186}{{\normalfont \ttfamily
  arXiv:hep-th/9503186}}.

\bibitem{Aharony:1995zh}
O.~Aharony, J.~Sonnenschein, M.~E. Peskin, and S.~Yankielowicz, ``{\em {Exotic
  nonsupersymmetric gauge dynamics from supersymmetric QCD}},''
  \href{http://dx.doi.org/10.1103/PhysRevD.52.6157}{Phys. Rev. D {\normalfont
  \bfseries 52} (1995)  6157--6174},
  \href{http://arxiv.org/abs/hep-th/9507013}{{\normalfont \ttfamily
  arXiv:hep-th/9507013}}.

\bibitem{Alvarez-Gaume:1996qoj}
L.~Alvarez-Gaume and M.~Marino, ``{\em {More on softly broken N=2 QCD}},''
  \href{http://dx.doi.org/10.1142/S0217751X97000724}{Int. J. Mod. Phys. A
  {\normalfont \bfseries 12} (1997)  975--1002},
  \href{http://arxiv.org/abs/hep-th/9606191}{{\normalfont \ttfamily
  arXiv:hep-th/9606191}}.

\bibitem{Alvarez-Gaume:1996vlf}
L.~Alvarez-Gaume, J.~Distler, C.~Kounnas, and M.~Marino, ``{\em {Softly broken
  N=2 QCD}},'' \href{http://dx.doi.org/10.1142/S0217751X96002170}{Int. J. Mod.
  Phys. A {\normalfont \bfseries 11} (1996)  4745--4777},
  \href{http://arxiv.org/abs/hep-th/9604004}{{\normalfont \ttfamily
  arXiv:hep-th/9604004}}.

\bibitem{DHoker:1996xdz}
E.~D'Hoker, Y.~Mimura, and N.~Sakai, ``{\em {Gauge symmetry breaking through
  soft masses in supersymmetric gauge theories}},''
  \href{http://dx.doi.org/10.1103/PhysRevD.54.7724}{Phys. Rev. D {\normalfont
  \bfseries 54} (1996)  7724--7740},
  \href{http://arxiv.org/abs/hep-th/9603206}{{\normalfont \ttfamily
  arXiv:hep-th/9603206}}.

\bibitem{Evans:1996hi}
N.~J. Evans, S.~D.~H. Hsu, and M.~Schwetz, ``{\em {Phase transitions in softly
  broken N=2 SQCD at nonzero theta angle}},''
  \href{http://dx.doi.org/10.1016/S0550-3213(96)00595-0}{Nucl. Phys. B
  {\normalfont \bfseries 484} (1997)  124--140},
  \href{http://arxiv.org/abs/hep-th/9608135}{{\normalfont \ttfamily
  arXiv:hep-th/9608135}}.

\bibitem{Konishi:1996iz}
K.~Konishi, ``{\em {Confinement, supersymmetry breaking and theta parameter
  dependence in the Seiberg-Witten model}},''
  \href{http://dx.doi.org/10.1016/S0370-2693(96)01527-4}{Phys. Lett.
  {\normalfont \bfseries B392} (1997)  101--105},
\href{http://arxiv.org/abs/hep-th/9609021}{{\normalfont \ttfamily
  arXiv:hep-th/9609021}}.

\bibitem{Alvarez-Gaume:1997bzm}
L.~Alvarez-Gaume, M.~Marino, and F.~Zamora, ``{\em {Softly broken N=2 QCD with
  massive quark hypermultiplets. 1.}},''
  \href{http://dx.doi.org/10.1142/S0217751X98000184}{Int. J. Mod. Phys. A
  {\normalfont \bfseries 13} (1998)  403--430},
  \href{http://arxiv.org/abs/hep-th/9703072}{{\normalfont \ttfamily
  arXiv:hep-th/9703072}}.

\bibitem{Alvarez-Gaume:1997wnu}
L.~Alvarez-Gaume, M.~Marino, and F.~Zamora, ``{\em {Softly broken N=2 QCD with
  massive quark hypermultiplets. 2.}},''
  \href{http://dx.doi.org/10.1142/S0217751X98000810}{Int. J. Mod. Phys. A
  {\normalfont \bfseries 13} (1998)  1847--1880},
  \href{http://arxiv.org/abs/hep-th/9707017}{{\normalfont \ttfamily
  arXiv:hep-th/9707017}}.

\bibitem{Evans:1997dz}
N.~J. Evans, S.~D.~H. Hsu, and M.~Schwetz, ``{\em {Controlled soft breaking of
  N=1 SQCD}},'' \href{http://dx.doi.org/10.1016/S0370-2693(97)00541-8}{Phys.
  Lett. B {\normalfont \bfseries 404} (1997)  77--82},
  \href{http://arxiv.org/abs/hep-th/9703197}{{\normalfont \ttfamily
  arXiv:hep-th/9703197}}.

\bibitem{Cheng:1998xg}
H.-C. Cheng and Y.~Shadmi, ``{\em {Duality in the presence of supersymmetry
  breaking}},'' \href{http://dx.doi.org/10.1016/S0550-3213(98)00539-2}{Nucl.
  Phys. B {\normalfont \bfseries 531} (1998)  125--150},
  \href{http://arxiv.org/abs/hep-th/9801146}{{\normalfont \ttfamily
  arXiv:hep-th/9801146}}.

\bibitem{Martin:1998yr}
S.~P. Martin and J.~D. Wells, ``{\em {Chiral symmetry breaking and effective
  Lagrangians for softly broken supersymmetric QCD}},''
  \href{http://dx.doi.org/10.1103/PhysRevD.58.115013}{Phys. Rev. D {\normalfont
  \bfseries 58} (1998)  115013},
  \href{http://arxiv.org/abs/hep-th/9801157}{{\normalfont \ttfamily
  arXiv:hep-th/9801157}}.

\bibitem{Luty:1999qc}
M.~A. Luty and R.~Rattazzi, ``{\em {Soft supersymmetry breaking in deformed
  moduli spaces, conformal theories, and N=2 Yang-Mills theory}},''
  \href{http://dx.doi.org/10.1088/1126-6708/1999/11/001}{JHEP {\normalfont
  \bfseries 11} (1999)  001},
  \href{http://arxiv.org/abs/hep-th/9908085}{{\normalfont \ttfamily
  arXiv:hep-th/9908085}}.

\bibitem{Cordova:2018acb}
C.~C\'ordova and T.~T. Dumitrescu, ``{\em {Candidate Phases for SU(2) Adjoint
  QCD$_4$ with Two Flavors from $\mathcal{N}=2$ Supersymmetric Yang-Mills
  Theory}},'' \href{http://arxiv.org/abs/1806.09592}{{\normalfont \ttfamily
  arXiv:1806.09592}}.

\bibitem{companion}
C.~Cs\'aki, A.~Gomes, H.~Murayama, and O.~Telem, ``{\em {The Phases of
  Non-supersymmetric Gauge Theories: the $SO(N)$ Case Study}},''
  forthcoming\!\!  .

\bibitem{Pomarol:1999ie}
A.~Pomarol and R.~Rattazzi, ``{\em {Sparticle masses from the superconformal
  anomaly}},'' \href{http://dx.doi.org/10.1088/1126-6708/1999/05/013}{JHEP
  {\normalfont \bfseries 05} (1999)  013},
  \href{http://arxiv.org/abs/hep-ph/9903448}{{\normalfont \ttfamily
  arXiv:hep-ph/9903448}}.

\bibitem{Intriligator:1995id}
K.~A. Intriligator and N.~Seiberg, ``{\em {Duality, monopoles, dyons,
  confinement and oblique confinement in supersymmetric SO(N(c)) gauge
  theories}},'' \href{http://dx.doi.org/10.1016/0550-3213(95)00159-P}{Nucl.
  Phys. B {\normalfont \bfseries 444} (1995)  125--160},
  \href{http://arxiv.org/abs/hep-th/9503179}{{\normalfont \ttfamily
  arXiv:hep-th/9503179}}.

\bibitem{Cardy:1981qy}
J.~L. Cardy and E.~Rabinovici, ``{\em {Phase Structure of Z(p) Models in the
  Presence of a Theta Parameter}},''
  \href{http://dx.doi.org/10.1016/0550-3213(82)90463-1}{Nucl. Phys. B
  {\normalfont \bfseries 205} (1982)  1--16}.

\bibitem{Cardy1982}
J.~L. Cardy, ``{\em Duality and the $\theta$ parameter in Abelian lattice
  models},''
  \href{http://dx.doi.org/https://doi.org/10.1016/0550-3213(82)90464-3}{Nuclear
  Physics B {\normalfont \bfseries 205} (1982) no.~1, 17--26}.

\bibitem{Seiberg:1994rs}
N.~Seiberg and E.~Witten, ``{\em {Electric - magnetic duality, monopole
  condensation, and confinement in N=2 supersymmetric Yang-Mills theory}},''
  \href{http://dx.doi.org/10.1016/0550-3213(94)90124-4}{Nucl. Phys. B
  {\normalfont \bfseries 426} (1994)  19--52},
  \href{http://arxiv.org/abs/hep-th/9407087}{{\normalfont \ttfamily
  arXiv:hep-th/9407087}}. [Erratum: Nucl.Phys.B 430, 485--486 (1994)].

\end{thebibliography}\endgroup

\end{document}